\documentclass[final,5p,times,twocolumn]{elsarticle}

\usepackage[hidelinks]{hyperref}

\usepackage{graphicx}
\usepackage{amsmath}   
\usepackage{array}
\usepackage{siunitx}
\usepackage[table,xcdraw]{xcolor} 

\begin{document}

\begin{frontmatter}

\title{4D-Tracking with Digital SiPMs}

	\author[1]{Inge Diehl}
	\author[1]{Finn Feindt}
    \author[1,2]{Ingrid-Maria Gregor}
	\author[1]{Karsten Hansen}
	\author[1,3]{Stephan Lachnit}
	\author[1,4]{Daniil Rastorguev}
	\author[1]{Simon Spannagel}
	\author[1]{Tomas Vanat}
	\author[1,2]{Gianpiero Vignola\corref{cor1}}
        \cortext[cor1]{Corresponding author}
        \ead{gianpiero.vignola@desy.de}
        
\affiliation[1]{organization={Deutsches Elektronen-Synchrotron DESY}, 
                 addressline={Notkestr. 85},
                 city={22607 Hamburg}, 
                 country={Germany}}
\affiliation[2]{organization={University of Bonn}, 
                 addressline={Regina-Pacis-Weg 3},
                 city={53113 Bonn},
                 country={Germany}}
\affiliation[3]{organization={University of Hamburg}, 
                 addressline={Mittelweg 177},
                 city={20148 Hamburg}, 
                 country={Germany}}
\affiliation[4]{organization={University of Wuppertal}, 
                 addressline={Gaußstraße 20},
                 city={42119 Wuppertal}, 
                 country={Germany}}

\begin{abstract}
 Silicon Photomultipliers (SiPMs) are the state-of-the-art technology in single-photon detection with solid-state detectors. Single Photon Avalanche Diodes (SPADs), the key element of SiPMs, can now be manufactured in CMOS processes, facilitating the integration of a SPAD array into custom monolithic ASICs. This allows implementing features such as signal digitization, masking, full hit-map readout, noise suppression, and photon counting in a monolithic CMOS chip. The complexity of the off-chip readout chain is thereby reduced. 

These new features allow new applications for digital SiPMs, such as 4D-tracking of charged particles, where spatial resolutions of the order of \SI{10}{\micro\meter} and timestamping with time resolutions of a few tens of ps are required.

A prototype of a digital SiPM was designed at DESY using the LFoundry \SI{150}{\nano\meter} CMOS technology. Various studies were carried out in the laboratory and at the DESY II test-beam facility to evaluate the sensor performance in Minimum Ionizing Particles (MIPs) detection. The direct detection of charged particles was investigated for bare prototypes and assemblies coupling dSiPMs and thin LYSO crystals. Spatial resolution \SI{\sim20}{\micro\meter} and a full-system time resolution of \SI{\sim50}{\pico\second} are measured using bare dSiPMs in direct MIP detection. Efficiency \SI{>99.5}{\percent}, low noise rate and time resolution \SI{<1}{\nano\second} can be reached with the thin radiator coupling. 

\end{abstract}

\end{frontmatter}

\section{Introduction}
\label{sec:introduction}
Silicon Photomultipliers (SiPMs) represent a solid-state version of conventional photomultipliers (PMTs). SIPMs are arrays of Single Photon Avalanche Diodes (SPADs) read out in parallel. SiPMs produce a signal approximately proportional to the number of incident photons, which is then processed with dedicated electronics. SiPMs are operated at relatively low voltages, have high gain, good photon sensitivity, excellent intrinsic timing, and are insensitive to magnetic fields. These characteristics fulfil the requirements of most High Energy Physics (HEP) photodetectors~\cite{Gundacker_2020}.

Some commercial CMOS foundries have introduced SPAD diodes in their Process Design Kits (PDKs). It is therefore possible to design digital SiPM integrated circuits (dSiPM from now on) where CMOS electronics and avalanche diodes are on the same silicon die. Signal digitisation can take place at the level of individual SPADs. Customised circuits and readout architectures can be implemented, simplifying the DAQ system and reducing the system power consumption. Low-cost implementation of custom designs and large-volume production are further advantages offered by CMOS foundries. 
The main drawbacks of CMOS dSiPMs are the smaller fill factor and higher Dark Count Rate (DCR) compared to conventional SiPMs~\cite{Bronzi2016SPADFO}. 

CMOS SPADs development is mainly driven by commercial applications such as Light Detection and Ranging (LIDAR) systems and 3D-Imaging~\cite{Lidar}. However, the properties of these sensors make them excellent candidates in new HEP applications such as scintillating fibres readout~\cite{fibers} or 4D-tracking of Minimum Ionizing Particles (MIPs), which is the main topic of this work.

The possibility of direct detection of MIPs with SPADs and SiPMs has already been proven~\cite{MIP_SPAD}~\cite{MIP_SiPM}. Due to the large number of electron-hole pairs created by an impinging MIP, an avalanche is triggered, with a high probability, and within few \SI{10}{\pico\second}, as soon as a SPAD is hit. Through the integration in a CMOS circuit, it is possible to develop chip readout architectures typical of tracking detectors with SPADs as MIP-sensitive pixels. 

\section{DESY dSiPM Prototype in CMOS 150 nm Technology}
\label{Sec:DESY_dSiPM}
\begin{figure}[tbp]
\centering
\includegraphics[width=\columnwidth]{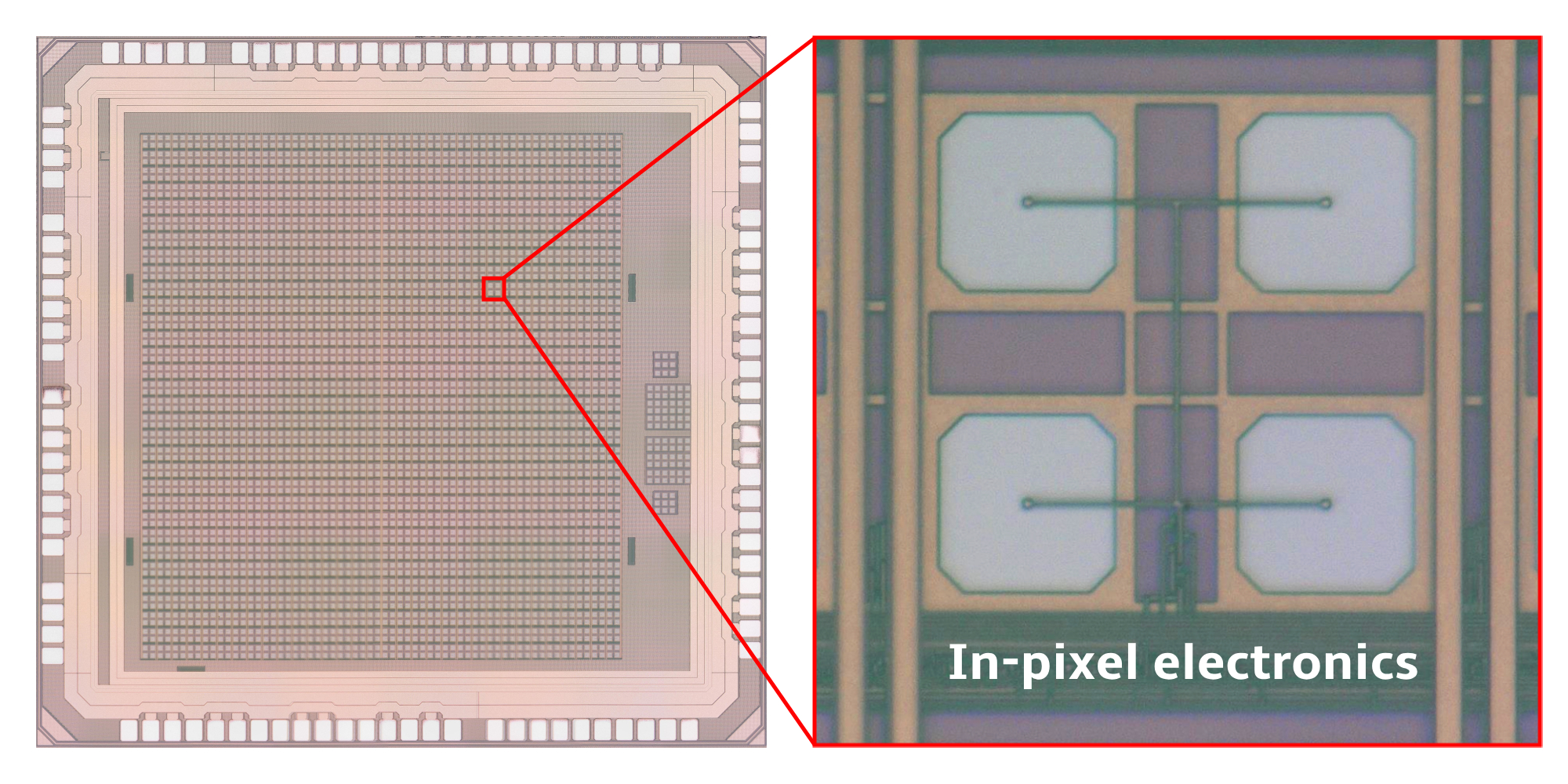}
\caption{Microscope picture of the DESY dSiPM chip (left). A magnified version of a pixel made of four SPADs sharing in-pixel electronics (right)}
\label{fig:Sensor}
\end{figure}

A prototype of dSiPM was designed at DESY using a LFoundry \SI{150}{\nano\meter} CMOS technology~\cite{Diehl_2024}. The chip includes pre-defined SPAD layout provided by the company and CMOS electronics. The main array includes \SI{32}\times\SI{32} pixels and has a total area of \SI{\sim 2.2}\times\SI{\sim2.4}{\milli\meter\squared}. Each pixel contains four SPADs connected in parallel that share the same in-pixel electronics. Figure~\ref{fig:Sensor} shows a microscope picture of a diced chip (left) and a zoom into a single pixel (right). The total pixel size is \SI{69.6}\times\SI{76}{\micro\meter\squared}. The four SPADs of the pixel have a nominal active area of \SI{20}\times\SI{20}{\micro\meter\squared} (white regions in Figure~\ref{fig:Sensor} right), the remaining area of the pixel is occupied by trenches and electronics well. This results in an effective fill factor of the pixel (i.e. of the sensor) of \SI{30}{\percent}. In-pixel electronics include a globally biased quenching transistor and a masking circuit.  An inverter for signal digitisation and a 2-bit hit counter are also part of the in-pixel circuit. In the chip periphery, four shared Time-to-Digital Converters (TDCs) allow time-stamping with \SI{\sim95}{\pico\second} bins. The full hit-map can be read out and validation logic is implemented for event discrimination. 

All chip characterizations was performed using the versatile Caribou DAQ system~\cite{Caribou}. It provides all bias and reference currents/voltages as well as clocks required for chip operations. The system is also used for chip configuration and data readout thanks to high-speed LVDS links and the system-embedded FPGA. The readout is frame-based and operates at \SI{3}{\mega\hertz} frame rate. More details on the sensor design and the DAQ system, together with chip characterizations can be found in~\cite{Diehl_2024}. 

\section{Test-Beam Characterization of DESY dSiPM}
\label{sec:TB}
In order to evaluate sensor performances in 4D-tracking of MIPs, several test-beam experiments were carried out on prototypes. Two main configurations have been investigated: bare silicon (Section~\ref{sec:bare_dSiPM}) and dSiPMs coupled with thin radiators (Section~\ref{sec:Thin_radiator} and Section~\ref{sec:dSiPM+LYSO}).

Test-beam studies took place at the DESY II test-beam facility~\cite{DESYII}. The facility provides electron beams in the energy range \SI{1}{} to \SI{6}{\giga\electronvolt} as well as reference beam telescopes for precise tracking~\cite{EUDET_Telescope}~\cite{ADENIUM}. An AIDA Trigger Logic Unit (TLU)~\cite{TLU} and the EUDAQ framework~\cite{EUDAQ} were employed for devices synchronisation and data acquisition. A fast scintillator was used to get a region of interest trigger and another scintillator with a hole as a veto. Active cooling allowed for stable operations at different temperatures down to \SI{0}{\celsius}. The precise temperature of the dSiPM was monitored using a temperature diode implemented in the chip periphery. 

\begin{figure}[tbp]
\centering
\includegraphics[width=\columnwidth]{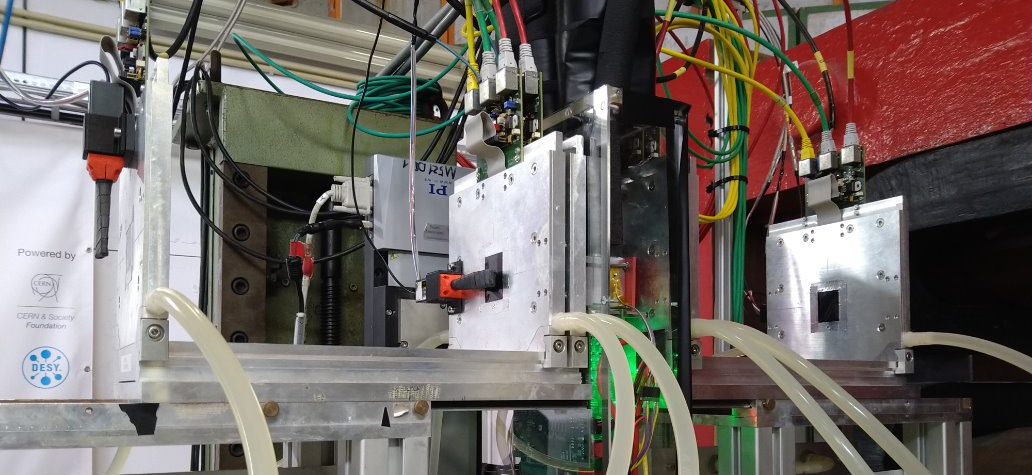}
\caption{Test-beam setup in the DESY II facility using EUDET type telescopes as reference for track reconstruction and two cooled DUTs in the middle.}
\label{fig:Setup}
\end{figure}

Figure~\ref{fig:Setup} shows the test-beam setup, a schematic sketch is shown in~\cite{Finn}, Figure~3. Two mechanically aligned Devices Under Test (DUTs) were investigated simultaneously to allow timing studies using time differences. Details of the setup and measurement campaigns can be found in~\cite{Finn}.

Test-beam data analysis was performed using the Corryvreckan framework~\cite{CORRY}.

\section{Performance of Bare dSiPM}
\label{sec:bare_dSiPM}
In the first test-beam campaigns bare dSiPMs were exposed to the electron beam to evaluate their spatial and temporal performance in direct MIP detection. The main results of these measurements are reported in~\cite{Finn} and the data analysis of further investigations is ongoing. 

\begin{figure}[tbp]
\centering
\includegraphics[width=\columnwidth]{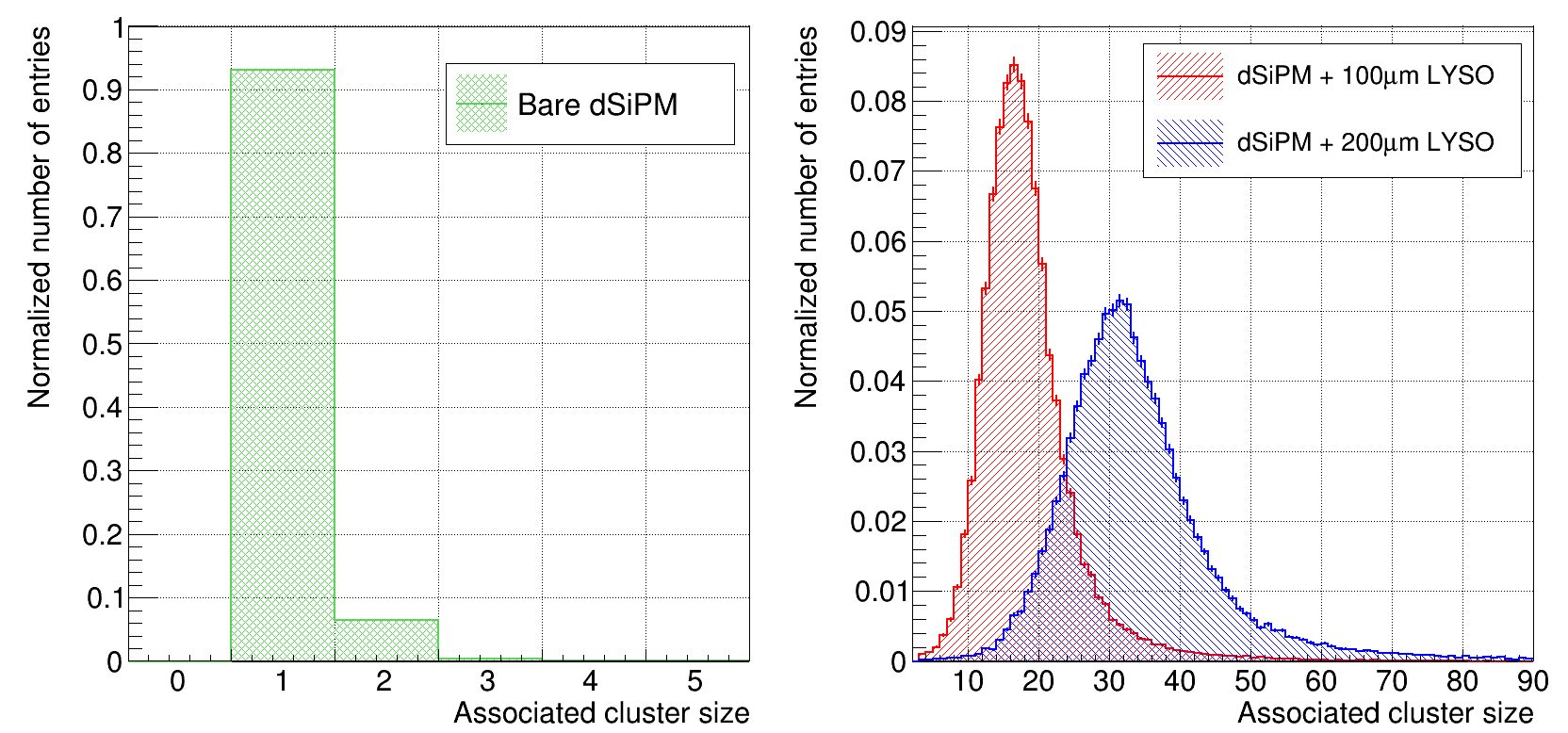}
\caption{Normalized associated cluster size of MIP events in test-beam for bare dSiPM (green), dSiPM+\SI{100}{\micro\meter} LYSO (red) and dSiPM+\SI{200}{\micro\meter} LYSO (blue). Plots refer to dSiPMs operated at \SI{0}{\celsius} and \SI{2}{\volt} overvoltage.}
\label{fig:Cluster_size}
\end{figure}

The cluster size of MIP events on bare dSiPM associated with reconstructed tracks is mainly one (see Figure~\ref{fig:Cluster_size} left). The cluster size is expressed in pixel units as the pixel readout structure does not provide information on the number of in-pixel firing SPADs. Cluster size larger than one is observed in a small percentage of the events and is consistent with the intrinsic crosstalk of SPADs. This means that noise hits can not be distinguished from MIPs, only based on event topology. In the test-beam environment though, the precise telescope tracking results in a few micrometer track position resolution at the DUT level allowing for the correct association of signal events with the corresponding track. 

\begin{figure}[tbp]
\centering
\includegraphics[width=\columnwidth]{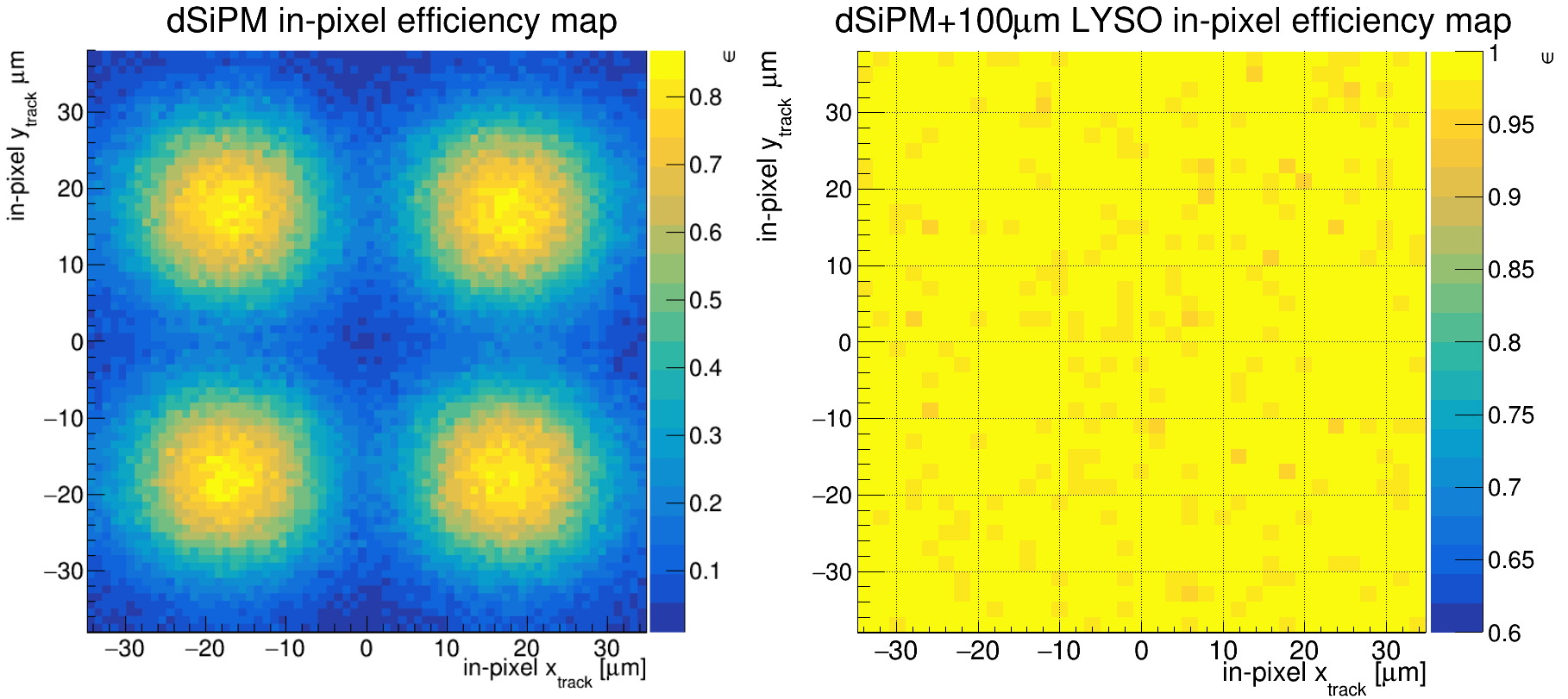}
\caption{In-pixel efficiency map in MIP detection for bare SiPM from~\cite{Finn} (left) and for dSiPM+\SI{100}{\micro\meter} LYSO (right). In the reported example, the samples are operated at \SI{0}{\celsius} and \SI{2}{\volt} overvoltage.}
\label{fig:Res&eff}
\end{figure}

In~\cite{Finn} spatial resolution in MIP detection is estimated from spatial residuals to be of the order of \SI{20}{\micro\meter} and efficiency is measured to be of the order of \SI{32}{\percent}. This value is comparable with the nominal fill factor of dSiPM. An in-pixel efficiency map can be obtained by projecting the particle intercepts with the DUT into a single pixel and averaging the efficiency for the binned in-pixel positions. Figure~\ref{fig:Res&eff} (left) from~\cite{Finn} shows an example of bare dSiPM in-pixel efficiency map. It can be noted that the pixel is efficient at SPADs and inefficient in the in-between area due to SPAD trenches and in-pixel electronics. 

The temporal resolution of the bare dSiPM in MIP detection is also investigated as described in~\cite{Lachnit:602203}. In the reference the time residuals are obtained by subtracting the timestamps of coincident MIP events in the two aligned DUTs. A temporal resolution of of 45~$\pm$~6~\SI{}{\pico\second} is measured in about \SI{85}{\percent} of the associated events. A slower component in nanosecond regime characterizes the remaining fraction of the data. This is due to MIP interactions in SPAD edge regions in one of the two DUTs. It should be emphasised that the timing performance reported here includes all contributions from the entire dSiPM DAQ system.

\section{The Thin Radiator Concept}
\label{sec:Thin_radiator}
One of the main drawbacks in direct MIP detection using bare dSiPMs is the limited efficiency due to the low fill factor. An inactive region between SPADs is needed to contain the avalanche within a single SPAD and additional space is used to host the in-pixel CMOS electronics. To overcome this limit a thin radiator coupled to the dSiPM can enhance the efficiency with a tolerable decrease of the spatial resolution.

In~\cite{MIP_SiPM} it was shown that using SiPM in direct MIP detection the resin employed to protect the sensor produces Cherenkov light. This improves the sensor efficiency and creates multi-SPAD signals. Following the same concept, scintillators with thicknesses of the O(\SI{100}{\micro\meter}) and high light yield can be coupled to the dSiPM. A MIP crossing the radiator releases energy and scintillation light is produced along its path. The light is isotropically emitted but due to the reduced thickness of the radiator is mostly confined and detected in a region close to the original MIP interaction point.

\begin{figure}[tbp]
\centering
\includegraphics[width=\columnwidth]{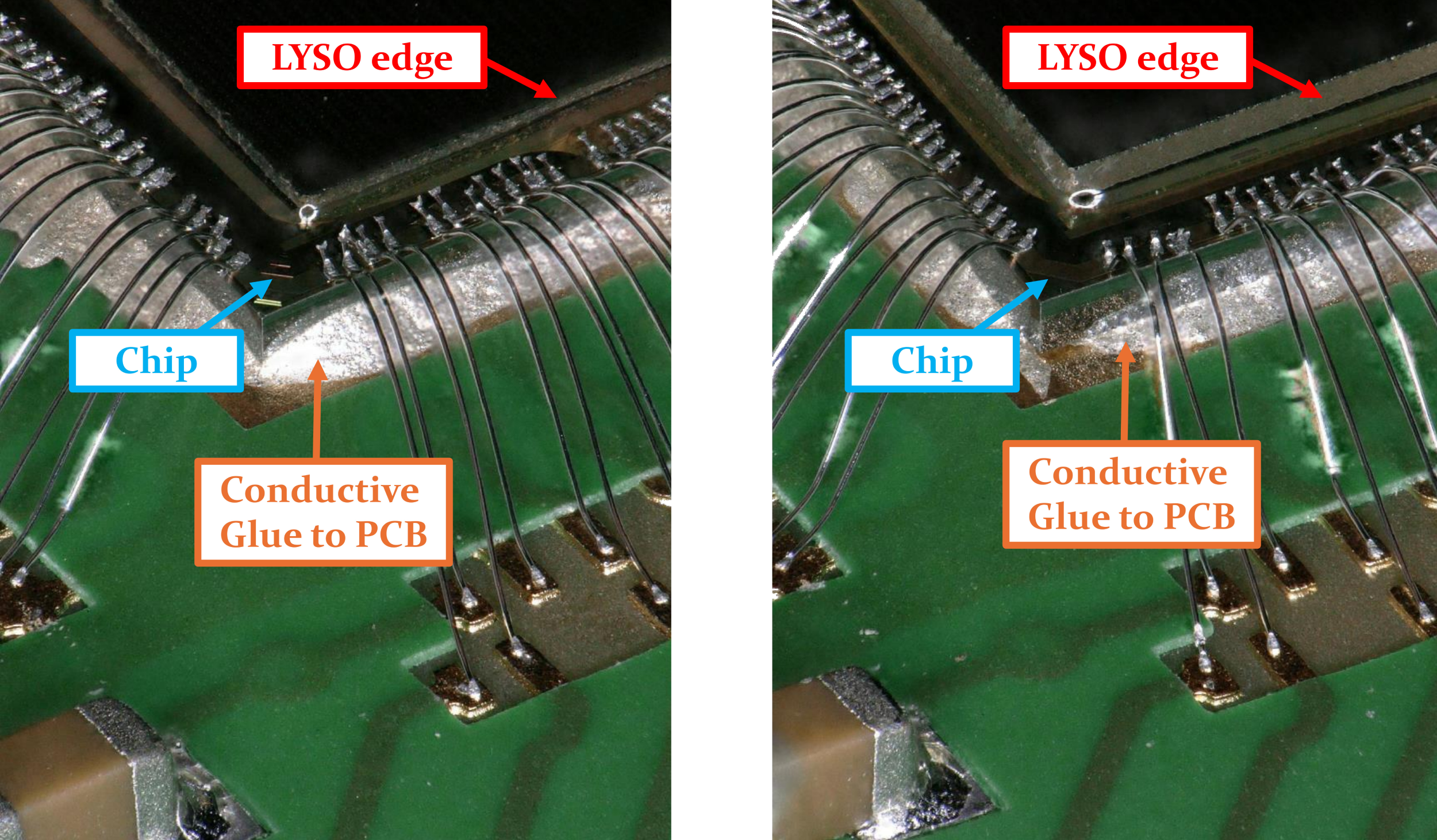}
\caption{Thin LYSO crystals glued on dSiPM samples. \SI{100}{\micro\meter} (left) and \SI{200}{\micro\meter} (right) thick LYSOs coupling is shown.}
\label{fig:LYSO}
\end{figure}

To test this concept, LYSOs with sizes of \SI{\sim 2.5}\times\SI{\sim2.5}{\milli\meter\squared} and thicknesses in the range \SI{100}{\micro\meter} to \SI{1}{\milli\meter} were produced by OST Photonics company with the two large faces polished. The crystals are coupled to dSiPM prototypes using Polytec EP 601-LV epoxy adhesive. Figure~\ref{fig:LYSO} shows two dSiPM samples coupled with \SI{100}{\micro\meter} (left) and \SI{200}{\micro\meter} (right) thick crystals.

\begin{figure}[tbp]
\centering
\includegraphics[width=\columnwidth]{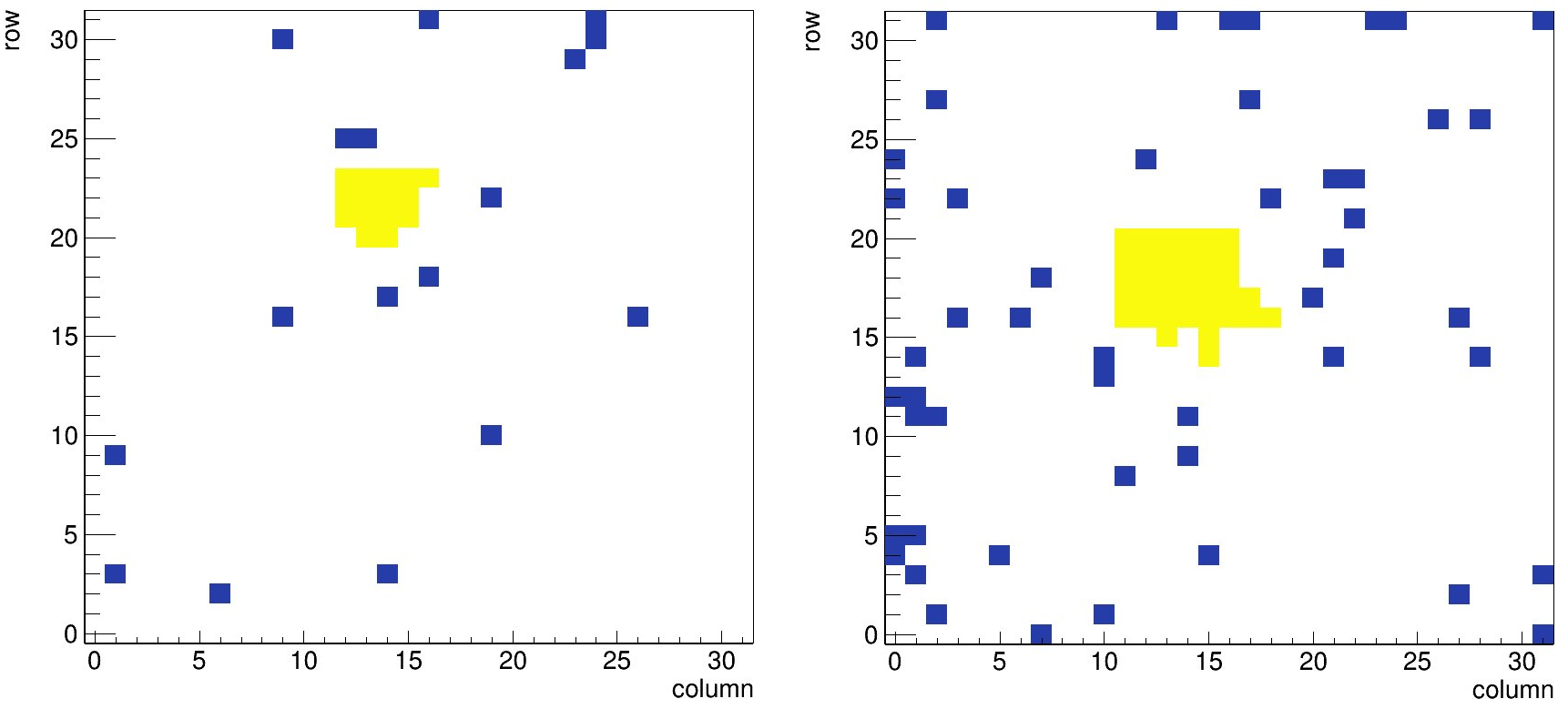}
\caption{Hit maps obtained with a Sr-90 source of dSiPM coupled with \SI{100}{\micro\meter} LYSO (left) and \SI{200}{\micro\meter} LYSO (right). Reconstructed clusters are marked in yellow. Chip operated at room temperature with \SI{2}{\volt} overvoltage}
\label{fig:Sr-90}
\end{figure}

After crystal coupling, the samples were tested using a Sr-90 source. Figure~\ref{fig:Sr-90} shows an example of a hit map obtained with a Sr-90 source for the two assemblies in Figure~\ref{fig:LYSO}. It can be noted that most of the hits are confined to clusters as expected. Only touching pixels are considered part of the cluster. The cluster centre is defined as the average position of pixels in the cluster. The original particle position can then be estimated as the cluster centre. 

\section{Performance of dSiPM with Thin Radiator}
\label{sec:dSiPM+LYSO}
Further test-beam campaigns were carried out on the prototypes coupled with thin radiators to evaluate the 4D-tracking performance. 

The first major difference in MIP detection with thin LYSOs coupling can be seen in the associated cluster size. Figure~\ref{fig:Cluster_size} shows the normalised cluster size of MIPs events for bare dSiPM, dSiPM+\SI{100}{\micro\meter} LYSO and dSiPM+\SI{200}{\micro\meter} LYSO. The comparison is made between runs where the chips were operated at the same temperature and overvoltage. The bare silicon events have a cluster size that is mainly one. With the thin LISO crystals, the cluster size increases considerably due to scintillation photons. This way, MIP events can be clearly distinguished from noise events. Noise signals involve dominantly one or two pixels (if crosstalk happens) and therefore have small cluster sizes (comparable with the bare dSiPM case). 
For geometrical reasons, the cluster size increases with the thickness of the radiator.  

The spatial resolution of the dSiPM+LYSOs prototypes was evaluated by analyzing at the spatial residuals between the track intercept with the DUT and the reconstructed cluster centre. The spatial resolution, defined as the sigma of the Gaussian fit on the spatial residuals, is \SI{\sim32}{\micro\meter}  for the \SI{100}{\micro\meter} LYSO sample and \SI{\sim38}{\micro\meter}  for the \SI{200}{\micro\meter} LYSO sample.

The MIP detection efficiency of dSiPMs coupled with thin LYSOs is defined as the ratio between the tracks intersecting the DUT and the subset of tracks with an associated cluster on the DUT. A spatial cut of \SI{200}{\micro\metre} and a time cut of \SI{2}{\micro\second} is applied in cluster association for this measurement. Efficiency is measured to be \SI{>99.5}{\percent} for all the tested overvoltages and temperatures. An example of an in-pixel efficiency map is shown in Figure~\ref{fig:Res&eff} (right). Here the pixel is fully efficient with no dead regions contrary to what is presented in Figure~\ref{fig:Res&eff} (left) for the bare dSiPM case, where only \SI{\sim 32}{\percent} of the pixel is efficient. 

The temporal resolution of dSiPM+LYSOs is currently being investigated. Preliminary results show a time resolution in the nanosecond regime when the MIPs do not hit a SPAD directly. In this scenario timing response is assumed to be dominated by the scintillation process. Faster or higher light-yield scintillators could result in better timing. 

\section{Conclusion \& Outlook}
\label{subsec1}

Digital SiPM integrated circuits represent an interesting technology for future detector systems in HEP and beyond. Low power budget, on-chip digitisation and data processing, excellent granularity, fast timing, large volume and low-cost production are some of the potential advantages offered by CMOS dSiPMs. 

As a possible application of the technology, MIPs 4D-tracking is studied. The sensor was successfully integrated into a test-beam setup and operated as a particle detector. DESY dSiPM prototype was tested as bare silicon and coupled with thin LYSO crystals. The spatial and timing performance of the prototypes were investigated. The main results of the measurements are summarized in Table~\ref{tab:tab}. 

\begin{table}[tbp] 
    \centering 
    \begin{tabular}{|>{\bfseries}l|c|c|} 
        \hline 
        \rowcolor{gray!30} 
        \bfseries  & \bfseries dSiPM & \bfseries dSiPM + LYSOs\\ 
        \hline 
        \bfseries MIP Cluster Size & $\sim$ 1 pixel & $\sim$ 20-40 pixels\\ 
        \hline 
        \bfseries Spatial Resolution & $\sim$ \SI{20}{\micro\meter} & $\sim$ 32-\SI{38}{\micro\meter}\\ 
        \hline 
        \bfseries Efficiency & \SI{\sim 32}{\percent} & \SI{>99.5}{\percent}\\ 
        \hline 
        \bfseries Time Resolution & \SI{50}{\pico\second}* & \SI{<1}{\nano\second}**\\ 
        \hline 
    \end{tabular}
    \begin{minipage}{\textwidth}
    \vspace{0.5em}
        \footnotesize
        \textbf{*} In the SPAD active area \hspace{0.5em} \textbf{**} Work in progress
    \end{minipage}
    \caption{DESY dSiPM 4D-tracking performance in MIP detection. Bare dSiPM (first column) and dSiPM with thin LYSO coupling (second column).}
    \label{tab:tab}
\end{table}

The proof of concept of the thin radiator coupling was successful, demonstrating full efficiency, good spatial resolution, and low noise rates.  Further investigations will determine whether the technology can meet all 4D-Tracking requirements. In particular, the use of faster radiators or higher-performance SPADs in different CMOS process nodes could allow high-demanding timing performance to be met. 

\section{Acknowledgments}
The authors would like to thank A. Venzmer, E. Wüstenhagen, D. Gorski and F. Poblotzki for test-board and mechanical case design, test setup, chip assembly as well as cooling system design and implementation.

Measurements leading to these results have been performed at the Test-Beam Facility at DESY Hamburg (Germany), a member of the Helmholtz Association (HGF).

\bibliographystyle{elsarticle-num-names} 
\bibliography{main}

\begin{thebibliography}{16}
\expandafter\ifx\csname natexlab\endcsname\relax\def\natexlab#1{#1}\fi
\providecommand{\url}[1]{\texttt{#1}}
\providecommand{\href}[2]{#2}
\providecommand{\path}[1]{#1}
\providecommand{\DOIprefix}{doi:}
\providecommand{\ArXivprefix}{arXiv:}
\providecommand{\URLprefix}{URL: }
\providecommand{\Pubmedprefix}{pmid:}
\providecommand{\doi}[1]{\href{http://dx.doi.org/#1}{\path{#1}}}
\providecommand{\Pubmed}[1]{\href{pmid:#1}{\path{#1}}}
\providecommand{\bibinfo}[2]{#2}
\ifx\xfnm\relax \def\xfnm[#1]{\unskip,\space#1}\fi
\bibitem[{Gundacker and Heering(2020)}]{Gundacker_2020}
\bibinfo{author}{S.~Gundacker}, \bibinfo{author}{A.~Heering},
\newblock \bibinfo{title}{The silicon photomultiplier: fundamentals and applications of a modern solid-state photon detector},
\newblock \bibinfo{journal}{Phys. Med. Biol.} \bibinfo{volume}{65} (\bibinfo{year}{2020}) \bibinfo{pages}{17TR01}. \DOIprefix\doi{10.1088/1361-6560/ab7b2d}.
\bibitem[{Bronzi and et~al.(2016)}]{Bronzi2016SPADFO}
\bibinfo{author}{D.~Bronzi}, \bibinfo{author}{et~al.},
\newblock \bibinfo{title}{{SPAD Figures of Merit for Photon-Counting, Photon-Timing, and Imaging Applications: A Review}},
\newblock \bibinfo{journal}{IEEE Sensors Journal} \bibinfo{volume}{16} (\bibinfo{year}{2016}) \bibinfo{pages}{3--12}. \DOIprefix\doi{https://doi.org/10.1109/JSEN.2015.2483565}.
\bibitem[{Qi and Zhang(2023)}]{Lidar}
\bibinfo{author}{F.~Qi}, \bibinfo{author}{P.~Zhang},
\newblock \bibinfo{title}{{High-resolution multi-spectral snapshot 3D imaging with a SPAD array camera}},
\newblock \bibinfo{journal}{Opt. Express} \bibinfo{volume}{31} (\bibinfo{year}{2023}) \bibinfo{pages}{30118--30129}. \DOIprefix\doi{10.1364/OE.492581}.
\bibitem[{Fischer et~al.(2022)Fischer, Zimmermann, and Maisano}]{fibers}
\bibinfo{author}{P.~Fischer}, \bibinfo{author}{R.~K. Zimmermann}, \bibinfo{author}{B.~Maisano},
\newblock \bibinfo{title}{{CMOS SPAD sensor chip for the readout of scintillating fibers}},
\newblock \bibinfo{journal}{Nucl. Instrum. Methods Phys. Res., Sect. A} \bibinfo{volume}{1040} (\bibinfo{year}{2022}) \bibinfo{pages}{167033}. \DOIprefix\doi{https://doi.org/10.1016/j.nima.2022.167033}.
\bibitem[{Gramuglia and et~al.(2023)}]{MIP_SPAD}
\bibinfo{author}{F.~Gramuglia}, \bibinfo{author}{et~al.},
\newblock \bibinfo{title}{{Direct MIP detection with sub-10 ps timing resolution Geiger-Mode APDs}},
\newblock \bibinfo{journal}{Nucl. Instrum. Methods Phys. Res., Sect. A} \bibinfo{volume}{1047} (\bibinfo{year}{2023}) \bibinfo{pages}{167813}. \DOIprefix\doi{https://doi.org/10.1016/j.nima.2022.167813}.
\bibitem[{Carnesecchi and et~al.(2023)}]{MIP_SiPM}
\bibinfo{author}{F.~Carnesecchi}, \bibinfo{author}{et~al.},
\newblock \bibinfo{title}{{Understanding the direct detection of charged particles with SiPMs}},
\newblock \bibinfo{journal}{Eur. Phys. J. Plus} \bibinfo{volume}{138} (\bibinfo{year}{2023}) \bibinfo{pages}{337}. \DOIprefix\doi{10.1140/epjp/s13360-023-03923-4}.
\bibitem[{Diehl and et~al.(2024)}]{Diehl_2024}
\bibinfo{author}{I.~Diehl}, \bibinfo{author}{et~al.},
\newblock \bibinfo{title}{{Monolithic MHz-frame rate digital SiPM-IC with sub-100 $ps$ precision and 70 $\mu m$ pixel pitch}},
\newblock \bibinfo{journal}{JINST} \bibinfo{volume}{19} (\bibinfo{year}{2024}) \bibinfo{pages}{P01020}. \DOIprefix\doi{10.1088/1748-0221/19/01/P01020}.
\bibitem[{Vanat(2020)}]{Caribou}
\bibinfo{author}{T.~Vanat},
\newblock \bibinfo{title}{{{Caribou – A versatile data acquisition system}}},
\newblock \bibinfo{journal}{PoS} \bibinfo{volume}{TWEPP2019} (\bibinfo{year}{2020}) \bibinfo{pages}{100}. \DOIprefix\doi{10.22323/1.370.0100}.
\bibitem[{Diener and et~al(2019)}]{DESYII}
\bibinfo{author}{R.~Diener}, \bibinfo{author}{et~al},
\newblock \bibinfo{title}{{The DESY II test beam facility}},
\newblock \bibinfo{journal}{Nucl. Instrum. Methods Phys. Res., Sect. A} \bibinfo{volume}{922} (\bibinfo{year}{2019}) \bibinfo{pages}{265--286}. \DOIprefix\doi{https://doi.org/10.1016/j.nima.2018.11.133}.
\bibitem[{Hendrik and et~al.(2016)}]{EUDET_Telescope}
\bibinfo{author}{J.~Hendrik}, \bibinfo{author}{et~al.},
\newblock \bibinfo{title}{{{Performance of the EUDET-type beam telescopes}}},
\newblock \bibinfo{journal}{EPJ Tech. Instrum.} \bibinfo{volume}{3} (\bibinfo{year}{2016}) \bibinfo{pages}{7}. \DOIprefix\doi{10.1140/epjti/s40485-016-0033-2}.
\bibitem[{Liu and et~al.(2023)}]{ADENIUM}
\bibinfo{author}{Y.~Liu}, \bibinfo{author}{et~al.},
\newblock \bibinfo{title}{{ADENIUM — A demonstrator for a next-generation beam telescope at DESY}},
\newblock \bibinfo{journal}{JINST} \bibinfo{volume}{18} (\bibinfo{year}{2023}) \bibinfo{pages}{P06025}. \DOIprefix\doi{10.1088/1748-0221/18/06/P06025}.
\bibitem[{Baesso et~al.(2019)Baesso, Cussans, and Goldstein}]{TLU}
\bibinfo{author}{P.~Baesso}, \bibinfo{author}{D.~Cussans}, \bibinfo{author}{J.~Goldstein},
\newblock \bibinfo{title}{{The AIDA-2020 TLU: a flexible trigger logic unit for test beam facilities}},
\newblock \bibinfo{journal}{JINST} \bibinfo{volume}{14} (\bibinfo{year}{2019}) \bibinfo{pages}{P09019}. \DOIprefix\doi{10.1088/1748-0221/14/09/P09019}.
\bibitem[{Liu and et~al.(2019)}]{EUDAQ}
\bibinfo{author}{Y.~Liu}, \bibinfo{author}{et~al.},
\newblock \bibinfo{title}{{EUDAQ2 A flexible data acquisition software framework for common test beams}},
\newblock \bibinfo{journal}{JINST} \bibinfo{volume}{14} (\bibinfo{year}{2019}) \bibinfo{pages}{P10033}. \DOIprefix\doi{10.1088/1748-0221/14/10/P10033}.
\bibitem[{Feindt and et~al.(2024)}]{Finn}
\bibinfo{author}{F.~Feindt}, \bibinfo{author}{et~al.},
\newblock \bibinfo{title}{{The DESY digital silicon photomultiplier: Device characteristics and first test-beam results}},
\newblock \bibinfo{journal}{Nucl. Instrum. Methods Phys. Res., Sect. A} \bibinfo{volume}{1064} (\bibinfo{year}{2024}) \bibinfo{pages}{169321}. \DOIprefix\doi{https://doi.org/10.1016/j.nima.2024.169321}.
\bibitem[{Dannheim and et~al.(2021)}]{CORRY}
\bibinfo{author}{D.~Dannheim}, \bibinfo{author}{et~al.},
\newblock \bibinfo{title}{{Corryvreckan: a modular 4D track reconstruction and analysis software for test beam data}},
\newblock \bibinfo{journal}{JINST} \bibinfo{volume}{16} (\bibinfo{year}{2021}) \bibinfo{pages}{P03008}. \DOIprefix\doi{10.1088/1748-0221/16/03/P03008}.
\bibitem[{Lachnit(2024)}]{Lachnit:602203}
\bibinfo{author}{S.~Lachnit}, \bibinfo{title}{{T}ime {R}esolution of a {F}ully-{I}ntegrated {D}igital {S}ilicon {P}hoto-{M}ultiplier}, \bibinfo{type}{Masterarbeit}, University of Hamburg, \bibinfo{year}{2024}. \DOIprefix\doi{10.3204/PUBDB-2024-00529}, \bibinfo{note}{masterarbeit, University of Hamburg, 2024}.

\end{thebibliography}

\end{document}